\documentclass[%
 aip,
 amsmath,amssymb,
 reprint,%
]{revtex4-1}

\usepackage{graphicx}
\usepackage{dcolumn}
\usepackage{bm}

\usepackage{comment}

\usepackage[utf8]{inputenc}
\usepackage[T1]{fontenc}
\usepackage{mathptmx}
\usepackage{etoolbox}
\usepackage{multirow}
\usepackage{xcolor}

\makeatletter
\def\@email#1#2{%
 \endgroup
 \patchcmd{\titleblock@produce}
  {\frontmatter@RRAPformat}
  {\frontmatter@RRAPformat{\produce@RRAP{*#1\href{mailto:#2}{#2}}}\frontmatter@RRAPformat}
  {}{}
}%
\makeatother
\begin{document}

\preprint{AIP/123-QED}

\title[RC-DDFT]{Nonequilibrium relaxation of soft responsive colloids}
\author{José López-Molina}
\affiliation{Department of Applied Physics, University de Granada, Campus Fuentenueva S/N, 18071 Granada, Spain}
\author{Sebastien Groh}%
\affiliation{Physikalisches Institut, Albert-Ludwigs-Universität Freiburg, Hermann-Herder Straße 3, D-79104 Freiburg, Germany}%

\author{Joachim Dzubiella$^*$}
\email{joachim.dzubiella@physik.uni-freiburg.de}
\affiliation{Physikalisches Institut, Albert-Ludwigs-Universität Freiburg, Hermann-Herder Straße 3, D-79104 Freiburg, Germany}
\affiliation{Cluster of Excellence livMatS @ FIT - Freiburg Center for Interactive Materials and Bioinspired Technologies, Albert-Ludwigs-Universität Freiburg, D-79110 Freiburg, Germany}

\author{Arturo Moncho-Jordá$^*$}
\email{moncho@ugr.es}
\affiliation{Department of Applied Physics, University de Granada, Campus Fuentenueva S/N, 18071 Granada, Spain}
\affiliation{Institute Carlos I for Theoretical and Computational Physics, University de Granada, Campus Fuentenueva S/N, 18071 Granada, Spain}

\date{\today}

\begin{abstract}

Stimuli-responsive macromolecules display large conformational changes during their dynamics, sometimes switching states, and are an integral property for the development of soft functional materials. Here, we introduce a mean-field dynamical density functional theory (DDFT) for a model of responsive colloids (RCs) to study the nonequilibrium dynamics of a colloidal dispersion in time-dependent external fields, with a focus on the coupling of translational and conformational dynamics during their relaxation. Specifically, we consider soft Gaussian particles with a bimodal size distribution between two confining walls with time-dependent (switching-on and off) external gravitational and osmotic fields. We find a rich relaxation behavior of the systems in excellent agreement with particle-based Brownian dynamics (BD) computer simulations. In particular, we find time-asymmetric relaxations of integrated observables (wall pressures, mean size, liquid center-of-mass) for activation/deactivation of external potentials, respectively, which are tuneable by the ratio of translational and conformational diffusion time scales. Our work paves thus the way for studying the nonequilibrium relaxation dynamics of complex soft matter with multiple degrees of freedom and hierarchical relaxations. 
\end{abstract}

\maketitle

\section{Introduction}

Responsive systems have garnered considerable interest in the realm of soft matter science due to their dynamic nature and the ability to adapt to external stimuli.~\cite{Stuart2010}  These systems, comprising responsive colloids and macromolecules, exhibit remarkable adaptability, controlling their properties—such as the internal conformation of the particle,~\cite{Onuchic1997,Wu1998,Cho2006,IDPswitching} size,~\cite {Denton2002,Urich2016,Brijitta2019,Baul2021, baul2021modulating, Lin2020} shape,~\cite{Meng2013,Lee2019,Brijitta2019,Lim2014,Lim2016,Lim2016a,Harrer2019, del2024numerical,elancheliyan2022role} charge density,~\cite{Brijitta2019,Weyer2018} electric dipole~\cite{Cao1993} and orientation,~\cite{calef1983smoluchowski,chandra1990collective,chandra1988role} among others - in response to environmental changes. Such responsiveness, originating from their internal degrees of freedom (DoFs), allows for a nuanced interaction with surrounding particles and external fields, leading to significant alterations in their internal and collective dynamical properties,\cite{Motornov2007,Motornov2011,Kalaitzidou2008,Stuart2010,Cao1993,Cheung2005,Zhou2008,Hong2010,Dupuis2014,Shin2015,Bimodal2021,huang2016microgels} even leading to multi-relaxation time scales.~\cite{garbin2015interactions}

An important system of RCs is the one for which the particle size (that could represent the radius of gyration of a linear polymer coil,~\cite{Denton2002,Vettorel2010} or the diameter of a microgel particle~\cite{Winkler,Moncho-Jorda2016,Karg2019,Brijitta2019,Rovigatti2019,Scotti,Scotti2019,Scotti2021}) is the internal property that couples to the center of mass translational degrees of freedom: particles not only move, but also are able to swell/shrink in response to external stimuli such as changes in the solvent pH, salt concentration and temperature.\cite{Murray1995, Saunders1999, fernandez2011microgel, zhou2015volume, Bochenek} In addition, particle concentration can also provoke the squeezing of the particles, leading to interpenetration, deformation and compression.~\cite{Nikolov2020, Gnan2019} Precise control over the interplay between size localization and dynamics is paramount for achieving targeted functionality at specific locations and rates. Notable examples are the local modulation of uptake and release kinetics in soft polymer-based nanocarriers, such as microgels for local control of catalysis~\cite{Roa2017,Roa2018,Kanduc2020} or drug release.~\cite{Hamidi2008,Saunders2009,Vinogradov2010,Moncho-Jorda2019,MonchoJorda2020a}

Our understanding of soft colloids has been significantly enhanced through developments in equilibrium Density Functional Theory (DFT), which has provided a solid framework for studying the structure and phase behavior of soft materials under external potentials. Classical DFT has been successfully extended to explicitly include varying particle sizes as dynamic variables, describing how microscopic interactions and external potentials affect the equilibrium properties of these systems and enabling the exploration of size-dependent phenomena within polydisperse systems.~\cite{Pagonabarraga2000,Pagonabarraga2004,Denton2002, pagonabarraga2001practical} Recently, we demonstrated explicitly for a model of responsive colloids (RC)s with size polydispersity~\cite{Urich2016,Baul2021, baul2021modulating, Lin2020} how to employ DFT functionals (in RC-DFT) to study and control the localization of the (size) property in space by external fields.~\cite{moncho2024external,moncho2023liquid}

However, many interesting behaviors of responsive systems occur out of equilibrium. The dynamical nature of these systems is not only a testament to their adaptability but also to their potential in real applications. In this context, dynamical density functional theory (DDFT) serves as a powerful tool for exploring non-equilibrium processes by modeling the time evolution of the particle density distribution, $\rho(\mathbf{r}, t)$, driven by diffusive, overdamped Brownian dynamics in the presence of external fields and particle interactions.~\cite{te2020classical,marconi1999dynamic,archer2004dynamical,royall2007nonequilibrium, Witti} The theory can also be modified to incorporate reactions or switching~\cite{Vrugt2020a,MonchoJorda2020,Moncho-Jorda2021,Moncho-Jorda2022} Crucially, it leverages the adiabatic approximation, assuming that the correlations in a non-equilibrium state are akin to those in equilibrium.~\cite{te2020classical} This theory adapts the equilibrium concepts of DFT to non-equilibrium scenarios, predicting how responsive systems evolve over time. In contrast to conventional polydisperse systems,~\cite{Pagonabarraga2000,Pagonabarraga2004} internal degrees of freedom are also able to change during the nonequilibrium process. A systematic study on the coupled dynamics of translation and internal dynamics in the relaxation of a colloidal system as well as the appropriate DDFT is still missing in literature. 

In this work, we present an extension of classical DDFT to responsive systems (denoted by RC-DDFT) that allows us to efficiently investigate the dynamical relaxation in the presence of coupled DoFs under applied external potentials. In particular, we consider soft responsive colloids (RC), for which the size of the particles ($\sigma$) changes in response to the interactions with the rest of particles or with an applied external potential. This external potential depends on the position and also the size of the particle, i.e. $u_\textmd{ext}(\mathbf{r},\sigma)$, making the colloidal system inhomogeneous in terms of position and size. In addition to the external potential, this system of responsive colloids also requires the knowledge of the free energy landscape for the particle size, $\psi(\sigma)=-kT\ln p(\sigma)$ (where $p(\sigma)$ represents the parent size-distribution of a single RC),  which acts as an additional external potential controlling the size fluctuations.~\cite{Lin2020} Here, we focus on a system formed by bistable particles for which $p(\sigma)$ is described by a generic Landau-like bimodal size distribution,~\cite{Tanaka1988,Suzuki1995} so particle size fluctuate between two states (big and small) separated by an energy barrier. This particular two-states behavior is relevant in the conformation of many biological or functional macromolecules, such as folded/unfolded or globule/coil transitions of proteins and polymers.~\cite{twostate1,twostate3,twostate2,IDP_switching,helical_switching,Dupuis2014,Kang2015} For this system, we analyse the time evolution of the one-body density profile, $\rho(\mathbf{r},\sigma;t)$, after sudden activation/deactivation of the external field, and explore the non-equilibrium transient dynamics resulting of the interplay between structural relaxation and size relaxation.

This paper is organized as follows. In Sec.~\ref{sec:theory} we describe the main statistical mechanics equations, discuss DFT and generalize its dynamical counterpart to deal with non-equilibrium systems of responsive colloids under external potentials (RC-DDFT). We also introduce in this Section a mean field model for soft Gaussian colloids with a bimodal distribution of states. Brownian dynamic simulations of RCs are explained in Sec.~\ref{sec:simulations}. Sec.~\ref{sec:resultsdiscussion} presents the results and discussion of the non-equilibrium dynamics of RCs under the activation/deactivation non-equilibrium processes for two representative external potentials: gravitation and osmotic. We specially focus on investigating the appearance of non-equilibrium transient dynamics states that arise due to the interplay between translational diffusion and particle swelling/shrinking. Finally, in Sec.~\ref{sec:conclusions} we present the main conclusions of our work.

\section{Theoretical background}
\label{sec:theory}

\subsection{\label{sec:theoryA}Theoretical modelling of responsive colloids (RCs)}

In the following, we briefly recall the most basic statistical relations between distributions and averages for the RC model.~\cite{Lin2020,MonchoJorda2023} As common in the theory of liquids, we assume a (isotropic) distance-dependent pair potential for the RC liquid. For RCs, however, the particle-particle pair potential not only depends on the positions $\mathbf{r}_i$ and $\mathbf{r}_j$ of both particles $i$ and $j$, but also has explicit dependence on the size of both interacting particles, $\sigma_i$ and  $\sigma_j$, that is $u(|\mathbf{r}_i-\mathbf{r}_j|;\sigma_i,\sigma_j)$. The external potential can be expressed as $u_\textmd{ext}(\mathbf{r}_i,\sigma_i)$. Hence, the total potential energy of $N$ responsive particles, can be expressed as
\begin{equation}
U = \sum_i \psi(\sigma_i) + \frac{1}{2}\sum_i\sum_{j\neq i} u(|\mathbf{r}_i-\mathbf{r}_j|,\sigma_i,\sigma_j) + \sum_i u_\textmd{ext}(\mathbf{r}_i,\sigma_i), 
\end{equation}
where $\psi$ is the energy landscape for the size $\sigma$ of an isolated particle. The inhomogeneous properties of an RC fluid immersed in a external potential are fully described by the particle density distribution, $\rho(\mathbf{r},\sigma)$,~\cite{Lin2020,MonchoJorda2023} defined in such a way that integration over the $4$ coordinates ($3$ translational ones and the size) provides the total number of particles in the system,
\begin{equation}
\label{norma}
\int_V d\mathbf{r}\int d\sigma \rho(\mathbf{r},\sigma)=N,
\end{equation}
so $\rho(\mathbf{r},\sigma)$ is measured in units of length$^{-4}$.

In the limit of very low particle concentrations and negligible external potential, the one-body density converges to $
\lim_{\rho_0\rightarrow 0}\ \lim_{u_\textmd{ext}\rightarrow 0}\rho(\mathbf{r},\sigma)=\rho_0p(\sigma)$, where $\rho_0=N/V$ is the bulk particle density, and $p(\sigma)$ is the so called \textit{parent} distribution, defined as the size distribution of a single isolated responsive particle that does not interact with external forces or with another particles, which is normalized to unity, $\int p(\sigma) d\sigma=1$. We can express $p(\sigma)$ in terms of the free energy landscape as
\begin{equation}
\label{ps}
p(\sigma)=p_0 e^{-\beta \psi(\sigma)},
\end{equation}
where $\beta=1/(k_BT)$ ($T$ is the absolute temperature and $k_B$ the Boltzmann constant), and $p_0$ is a constant prefactor to fulfill the normalization of $p(\sigma)$. For higher particle concentrations or in the presence of external fields, the single-particle property distribution will change: We denote by $f(\sigma)=N(\sigma)/N$ the {\it emergent} probability distribution, where $N(\sigma)d\sigma$ is the number of particles with internal property within $\left[\sigma,\sigma + d\sigma \right]$ in the system, and $N$ the total number of particles. In particular, in the absence of any external field, the one-body particle density may be expressed as $\lim_{u_\textmd{ext}\rightarrow 0}\rho(\mathbf{r},\sigma)=\rho_0f(\sigma)$. 
\subsection{Dynamical density functional theory for RCs}

\subsubsection{\label{sec:theoryB} Equilibrium DFT prerequisites}

Let us recall first the treatment of the RC model in the framework of equilibrium DFT. The inhomogeneous free energy functional of a RC fluid immersed in the external potential $u_\textmd{ext}(\mathrm{r},\sigma)$ is given by~\cite{Pagonabarraga2000,Lin2020,MonchoJorda2023}
\begin{eqnarray}
\label{F}
F [ \rho(\mathbf{r},\sigma)]&=&  k_BT\int d\mathbf{r}\int d\sigma \rho(\mathbf{r},\sigma) \big[ \ln (\rho(\mathbf{r},\sigma)\Lambda^3/p_0) -1  \big]    \nonumber \\
&+&\int d\mathbf{r}\int d\sigma \rho(\mathbf{r},\sigma)[u_\textmd{ext}(\mathbf{r},\sigma)+\psi(\sigma)]   \\
&+& F_\textmd{ex}[ \rho(\mathbf{r},\sigma)],  \nonumber
\end{eqnarray}
where $\Lambda=h/(2\pi m k_B T)^{1/2}$ is the thermal wavelength. The first term of Eq.~(\ref{F}) is the ideal gas free-energy functional. The second term takes into account the interaction of the RC fluid with the external potential. Note that $\psi(\sigma)$ also plays the role of an external potential for the particle size, i.e., it represents the energy cost implied in the swelling/shrinking of each responsive colloid. Finally, the third contribution is the excess free energy of the fluid that arises due to the existence of particle-particle interactions.

The grand canonical potential energy functional of a RC fluid is
\begin{equation}
\label{Omega}
\Omega [ \rho(\mathbf{r},\sigma)]= F[\rho(\mathbf{r},\sigma)]-  \mu_0\int d\mathbf{r}\int d\sigma \rho(\mathbf{r},\sigma),
\end{equation}
where $\mu_0$ is the (constant) chemical potential of the RC fluid. The equilibrium density profile is the one that minimizes the grand canonical functional, $\delta \Omega /\delta \rho(\mathbf{r},\sigma)=0$. Applying this functional differentiation to Eq.~(\ref{Omega}) with Eq.~(\ref{F}) and solving the resulting Euler-Lagrange equation for the particle density $\rho(\mathbf{r},\sigma)$, we find
\begin{equation}
\label{rhoite1}
\rho(\mathbf{r},\sigma)=q\exp \big(-\beta \psi(\sigma)-\beta u_\textmd{ext}(\mathbf{r},\sigma)-\beta \mu_\textmd{ex}(\mathbf{r},\sigma)\big),
\end{equation}
where $\mu_\textmd{ex}(\mathbf{r},\sigma)$ is the excess chemical potential, given by the functional differentiation of the excess free energy, $\mu_\textmd{ex}(\mathbf{r},\sigma)=\delta F_\textmd{ex}/\delta \rho(\mathbf{r},\sigma)$, and $q$ is a normalization constant that is obtained imposing conservation of the total number of particles, namely Eq.~(\ref{norma}). Solutions of Eqs.~(\ref{rhoite1}) and (\ref{norma}) lead to the equilibrium position and size distribution $\rho_\textmd{eq}(\mathbf{r},\sigma)$.

\subsubsection{Dynamical DFT for a RC fluid}

In non-equilibrium conditions, the one-body density distribution becomes also time-dependent, $\rho (\mathbf{r},\sigma;t)$. For the case of responsive colloids, particles in the system do not only diffuse in the space, but can also can modify their their size in response to external interactions. To describe their dynamics we make the assumption that the property $\sigma$ of each particle also follows an overdamped diffusive dynamics. Following the prescription given in the work by Baul et al.,~\cite{Baul2021} we denote $D_\textmd{T}$ as the translational diffusion coefficient, and $D_\sigma$ the diffusion coefficient in the $\sigma$-space. It is important to emphasize that $D_\textmd{T}$ depends in general on the particle size (as for Stokes-Einstein), so $D_\textmd{T}(\sigma)$ is a function of $\sigma$.

The DDFT can be extended to RCs by defining a 4-dimensional vector $\mathbf{x}=(x,y,z,\sigma)\equiv (\mathbf{r},\sigma)$, and a $4$-dimensional current $\mathbf{J}=(J_x,J_y,J_z,J_\sigma)\equiv(\mathbf{J}_{\mathbf{r}},J_\sigma)$ (note that this current has dimensions of length$^{-3}$time$^{-1}$). Analogously, we can write a $4$-component \textit{nabla} operator $\nabla_\mathbf{x}=(\partial/\partial x, \partial/\partial y, \partial/\partial z, \partial/\partial \sigma)\equiv (\nabla_{\mathbf{r}},\nabla_{\sigma})$. The time evolution of the density profile is given by
\begin{equation}
\label{RCDDFT}
\frac{\partial \rho(\mathbf{x},t)}{\partial t}=-\nabla_{\mathbf{x}}\cdot \mathbf{J}(\mathbf{x},t)=-\nabla_{r}\cdot \mathbf{J}_\mathbf{r}-\frac{\partial J_\sigma}{\partial \sigma}
\end{equation}

The components of the current $\mathbf{J}$ are
\begin{equation}
\begin{cases}
\mathbf{J}_\mathbf{r}=-D_\textmd{T}(\sigma)\rho(\mathbf{r},\sigma;t)\nabla_\mathbf{r}[\beta \mu(\mathbf{r},\sigma;t)] \\
J_\sigma=-D_\sigma\rho(\mathbf{r},\sigma;t)\frac{\partial \beta \mu(\mathbf{r},\sigma;t)}{\partial \sigma}
\end{cases}
\label{Jx}
\end{equation}
where $\mu(\mathbf{r},\sigma;t)$ is the non-equilibrium chemical potential. In order to calculate it, we make use of the adiabatic approximation, and assume that $\mu(\mathbf{r},\sigma;t)$ is given by the functional derivative of the equilibrium free energy functional, $\mu=\delta F/\delta \rho$. Using Eq.~(\ref{F}), it leads to~\cite{moncho2024external}
\begin{eqnarray}
\label{mumuex}
\mu(\mathbf{r},\sigma;t)&=&k_BT\ln \big( \rho(\mathbf{r},\sigma;t) \Lambda^3/p_0\big)+u_\textmd{ext}(\mathbf{r},\sigma)+\psi(\sigma) \nonumber \\
&+&\mu_\textmd{ex}(\mathbf{r},\sigma;t),
\label{eq:DDFT_non-pot}
\end{eqnarray}
where $\mu_\textmd{ex}=\delta F_\textmd{ex} / \delta \rho$. Eqs.~(\ref{RCDDFT})-(\ref{mumuex}) define the new generalized dynamical density functional theory designed for responsive colloids (RC-DDFT).

\subsection{\label{sec:theoryC}Mean-field responsive DDFT of soft Gaussian responsive colloids confined in planar slits}

In this section, we specify our particular system of responsive colloids, and the corresponding excess free energy model for it to be used in the RC-DDFT framework. In this work we focus on systems composed by soft interpenetrable RCs. We consider the following size-dependent Gaussian-core pair potential for the particle-particle interaction
\begin{equation}
	\label{u}
	\beta u(|\mathbf{r}-\mathbf{r}^{\prime}|,\sigma,\sigma^{\prime})=\epsilon_{ij} \exp \Big( -4|\mathbf{r}-\mathbf{r}^\prime|^2/(\sigma+\sigma^{\prime})^2 \Big),
\end{equation}
where $\epsilon>0$ is the (repulsive) interaction strength (in $k_BT$ units). It represents an estimate of the particle hardness: for small values of $\epsilon$ colloids are able to interpenetrate each other. On the contrary, large values of $\epsilon$ correspond to harder colloids, for which the energy penalty of overlapping is very high. Here, $(\sigma+\sigma^{\prime})/2$ plays the role of the interaction range. In fact, in this interaction model $\sigma$ represents the effective radius of the each RC.

The Gaussian-core interaction model is a well established coarse-grained description of polymer
solutions,~\cite{Likos2001a} as it has been shown to accurately describe the interaction between two isolated
polymers immersed in a good solvent, for polymer of identical~\cite{Bolhuis2001} and different length~\cite{Dautenhahn1994}, both in homogeneous and the inhomogeneous
conditions.~\cite{Louis2000} For different choices of the interaction parameters one can obtain either a mixture exhibiting
bulk fluid–fluid (macro)phase separation~\cite{Finken2000, Archer2001,Archer2005,Archer2004} similarly to polymer blends, or alternatively, a mixture exhibiting
microphase separation.~\cite{Archer2004b}

We assume that $\epsilon=2$ (in $k_BT$ units) for the interparticle interaction strength. This value represents fairly well the soft repulsion existing between linear polymers.~\cite{Louis2000,Archer2001} The equilibrium properties of such soft Gaussian particles described by Eq.~(\ref{u}) are well represented by a weakly correlated mean-field fluid over a surprisingly wide density and temperature range.~\cite{Louis2000} This justifies the use of the mean-field approximation for the excess free-energy of the interacting RC fluid, given by
\begin{equation}
\label{Fex}
F_\textmd{ex}=\frac{1}{2}\iint_V d\mathbf{r}d\mathbf{r}^{\prime}\iint d\sigma d\sigma^{\prime} \rho(\mathbf{r},\sigma)\rho(\mathbf{r}^{\prime},\sigma^{\prime})u(|\mathbf{r}-\mathbf{r}^{\prime}|,\sigma,\sigma^{\prime})
\end{equation}

This approximation has been successfully used to reproduce the  equilibrium and non-equilibrium properties of passive~\cite{Archer2005,Archer2005a} and active switching Gaussian colloids.~\cite{MonchoJorda2020,Moncho-Jorda2021,Moncho-Jorda2021,Moncho-Jorda2022}

Performing the functional differentiation and introducing it into Eq.~(\ref{eq:DDFT_non-pot}) leads to the explicit expression for the non-equilibrium chemical potential of a mean-field RC fluid
\begin{eqnarray}
\label{mumf}
    \mu(\mathbf{r},\sigma;t)&=&k_BT\ln \big( \rho(\mathbf{r},\sigma;t) \Lambda^3/p_0\big)+u_\textmd{ext}(\mathbf{r},\sigma)+\psi(\sigma) \nonumber \\
    &+&\int d\mathbf{r}^\prime \int d\sigma^\prime \rho(\mathbf{r}^{\prime},\sigma^{\prime};t)u(|\mathbf{r}-\mathbf{r}^{\prime}|,\sigma,\sigma^{\prime}),
    \label{eq:DDFT_chemical}
\end{eqnarray}
which involves a convolution integral in the $\mathbf{r}$ coordinate. 

In our work,  we consider RC dispersions confined between two parallel hard walls separated by a distance $L$ and subjected to one-dimensional external potentials, $u_\textmd{ext}(z,\sigma)$, where $z \in [0, L]$ is the distance from the left wall. The rest of coordinates are assumed to run over the full range, $x,y \in ]-\infty,\infty [$ (infinite slit). In this case the density profiles are homogeneous in lateral directions and can be expressed as $\rho(\mathbf{r},\sigma;t)=\rho(z,\sigma;t)$, with the normalization
\begin{equation}
	\int_0^Ldz\int d\sigma \rho(z,\sigma; t)=N/S, 
    \label{eq:density}
\end{equation}
where $N/S$ is the prefixed number density per area $S$, and the normalization is valid for every time $t$.

Exploiting the planar symmetry to simplify the convolution integral involved in Eq.~(\ref{eq:DDFT_chemical}), we find the following equation for the non-equilibrium chemical potential~\cite{moncho2024external}
\begin{eqnarray}
\label{eq:DDFT_chem_slit}
\mu(z,\sigma;t)&=k_BT\ln \big( \rho(z,\sigma;t) \Lambda^3/p_0\big)+u_{ext}(z,\sigma)+\psi(\sigma) \nonumber \\
+&\frac{\pi\epsilon}{4}\int d\sigma^{\prime}(\sigma+\sigma^\prime)^2\int_0^L dz^\prime \rho(z^{\prime},\sigma^{\prime};t)e^{-\frac{4(z-z^\prime)^2}{(\sigma+\sigma^\prime)^2}}.
\end{eqnarray}

In addition to the interparticle interaction potential, we need to specify the parent distribution of the RC, $p(\sigma)$. In this work we consider bistable particles, for which the size fluctuates between two states of size $\sigma_1$ and $\sigma_2$, that can be referred as small and big. This implies considering a bimodal parent size distribution, with both peaks centered around these two states. To model this behavior, we choose a generic bimodal form using a symmetric double-Gaussian function
\begin{equation}
\label{eq:bimodalsim}
p(\sigma) = \frac{p_0}{2\sqrt{2\pi \tau^2}}\Big[ \exp \Big( -\frac{(\sigma-\sigma_1)^2}{2\tau^2}\Big) + \exp\Big( -\frac{(\sigma-\sigma_2)^2}{2\tau^2}\Big) \Big]
\end{equation} 
with $\sigma_1=0.63\sigma_0$, $\sigma_2=1.37\sigma_0$ ($\sigma_0$ represents a reference particle size that will be used as unit length for the rest of sizes and distances). In order to avoid nonphysical negative values and extremely large values of the particle size, the range of $\sigma$ has been limited to be $\sigma \in [0.1\sigma_0,2\sigma_0]$. 

The parameter $\tau$ appearing in Eq.~(\ref{eq:bimodalsim}) provides the thickness of the size distribution around both peaks, and can be interpreted as the particle softness (conversely, $\tau^{-1}$ represents the stiffness of the RC). In this work we use $\tau=0.2\sigma_0$. With this choice, the free energy barrier required to overcome to switch from one state to other is $\Delta \psi \approx 1 k_BT$.

In our system, the $\sigma$-dependence of the translational diffusion coefficient will be of the type of Stokes, $D_\textmd{T}(\sigma)=D_0\sigma_0/\sigma$, where $D_0$ is the diffusion coefficient of a particle of radius $\sigma_0$. In addition, a parameter $\alpha$ is introduced to control the ratio between $\sigma$-diffusion and translational diffusion, $D_\sigma=\alpha D_0$. We also define our diffusion time for either translation or size relaxation as $\tau_{\rm B}=\sigma_0^2/D_0$.

From $\rho(z,\sigma;t)$ and integrating over the $\sigma$-coordinate we obtain the one-body number density distribution of the RC fluid in our quasi-1D system, namely
\begin{equation}
\label{eq:rhor}
\rho(z;t) = \int d\sigma \rho(z,\sigma;t).
\end{equation}

We analyze in our work a few integrated properties and monitor their time evolution: The center of mass location of the RC fluid is given by
\begin{equation}
\label{eq:meanz}
\langle z(t) \rangle = \frac{S}{N}\int_0^L zdz\int d\sigma \rho(z,\sigma;t).
\end{equation}

The mean size of the RC at position $z$ is given by the normalized first moment of the distribution 
\begin{equation}
	\label{eq:sigmar}
	\langle \sigma(z;t) \rangle=\frac{1}{\rho(z;t)}\int d\sigma \rho(z,\sigma;t)\sigma.
\end{equation}

The (global) mean size of the RC is given by

\begin{equation}
	\label{eq:meansigma}
	\langle \sigma(t) \rangle= \frac{S}{N}\int_0^L dz\int d\sigma \rho(z,\sigma;t)\sigma.
\end{equation}

The pressure exerted on the left and right wall are given respectively by

\begin{equation}
	\label{eq:pressleft}
	P_{\rm left}(t)= \rho(0;t)k_BT \ \ {\rm and} \ \ P_{\rm right}(t)= \rho(L;t)k_BT.
\end{equation}

\section{Brownian Dynamics simulations of RCs}
\label{sec:simulations}

Brownian dynamics (BD) calculations are performed in the framework of the RC model.\cite{Lin2020} The discretized form of the BD equations for the translational degrees of freedom and for the internal property are given by:
\begin{equation}
\begin{cases}
    \mathbf{r}_i(t+\Delta t) = \mathbf{r}_i(t) + \frac{D_\textmd{T}}{k_BT}\mathbf{F}_\textmd{T}^{(i)}(t) \Delta t + {\bm \xi}_\textmd{T} \\
    \sigma_i (t+\Delta t) = \sigma_i(t) + \frac{D_{\sigma}}{k_BT} F_{\sigma}^{(i)}(t) \Delta t + \xi_{\sigma}
\end{cases}
\end{equation}
where $\Delta t$ is the simulation time step. While $\mathbf{F}_\textmd{T}^{(i)}(t) = -\bm \nabla_{\bm r_i} u_\textmd{ext}(\sigma_i, z; t) - \sum_{j\ne i}^N \bm \nabla_{\bm r_i} u(\mathbf{r}_i-\mathbf{r}_j, \sigma_i, \sigma_j)$ is the translational force acting on colloid $i$ from both the external field and the pairwise interactions with the other colloids,  $F_{\sigma}^{(i)}(t) = -\nabla_{\sigma_i} \psi(\sigma_i) - \nabla_{\sigma_i} u_\textmd{ext}(\sigma_i, z; t) - \sum_{j \ne i}^N \nabla_{\sigma_i} u(\mathbf{r}_i-\mathbf{r}_j, \sigma_i, \sigma_j)$ is the property force acting on colloid $i$ resulting from its own free-energy landscape $\psi(\sigma)$, and its interaction with both the external field and the pairwise interactions with the other colloids. Finally, $\bm \xi_T$ and  $\xi_{\sigma}$ are random vector and random scalar, respectively. Both $\bm \xi_T$ and  $\xi_{\sigma}$ are drawn from a normal distribution with zero mean and variance $\langle \xi_{T,\alpha}\xi_{T,\beta}\rangle = 2D_\textmd{T} \Delta t \delta_{\alpha \beta}$ and $\langle \xi^2_{\sigma}\rangle = 2D_{\sigma}\Delta t$, respectively, with $\alpha$, $\beta = x, y, z$ being the Cartesian components, and $\delta_{\alpha \beta}$ being the Kronecker delta function. 

Simulations were performed using the same geometry and potential parameters as the ones used in RC-DDFT. To avoid non vanishing, non-physical negative values, and extremely large values of the colloid size, the range of $\sigma$ was limited to $[0.1,2]\sigma_0$ by simply rejecting (Monte-Carlo like) moves that would lead to $\sigma < 0.1\sigma_0$ or $\sigma > 2 \sigma_0$. The same procedure was used to respect the hard walls. Starting with an equilibrium configuration made of 432 responsive colloids (see Fig.~\ref{fig:fig1}(a)), the system was perturbed by either switching on or off $\phi_\textmd{G}$ or $\phi_\textmd{O}$, prior to a relaxation run of $30\tau_{\text{B}}$ during which the system reached a new equilibrium state (see Fig.~\ref{fig:fig1}(c)). To obtain the time-dependent density, $\rho(z;t)$, and mean size, $\langle \sigma (t)\rangle$, $N_{run} = 4000$ independent runs were performed. The density profile and the mean size were then averaged over all configurations for a fixed time $t$.


\section{Results and discussion}
\label{sec:resultsdiscussion}

For our non-equilibrium relaxation study, our methodology consists on applying an external potential, 
\begin{equation}
u_{\text{ext}}(z;\sigma;t) = u_{\text{walls}}(z) + \phi(z,\sigma;t),
\end{equation}
where $u_{\text{walls}}$ accounts for the confinement effects, and $\phi$ represents additional external potentials such as osmotic or gravitational potentials, defined further below. In particular, the wall potential is 

\begin{equation}
	u_{\text{walls}}(z)=\begin{cases}
		0 \quad \ \text{for} \quad 0 \le z \le L  \\
		\infty \quad \text{for} \quad  z < 0 \ \ \textmd{or} \ \ z > L 
	\end{cases}
	\label{eq:u_wall}\nonumber  
\end{equation}

\begin{figure}[ht!]
  \centering
    \includegraphics[width=\linewidth]{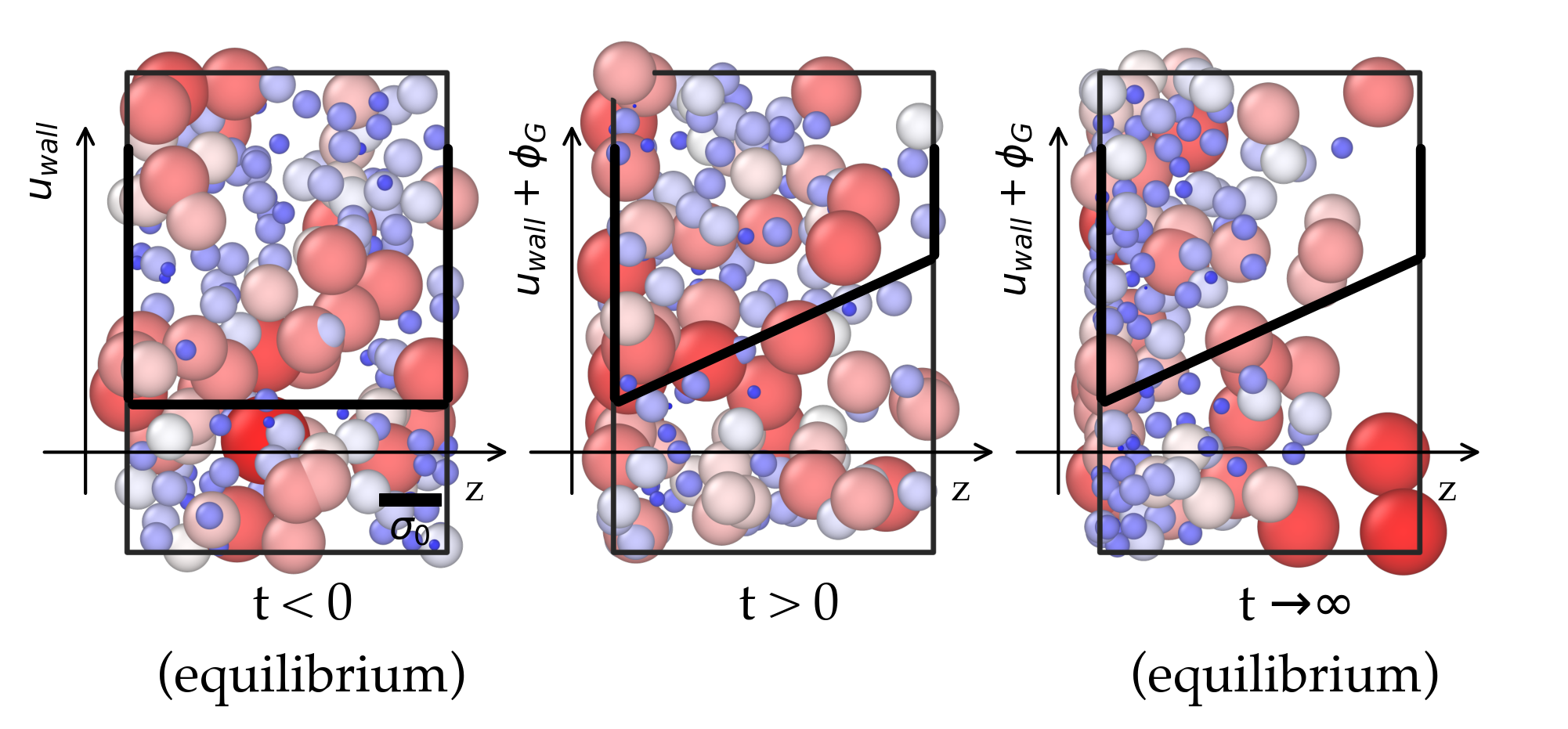}
\caption{Representative snapshots of system configurations made of responsive colloids obtained at different times from Brownian dynamics simulations. For $t < 0$, the system is in equilibrium with hard walls on the left and on the right (solid line). At $t=0$, the system is perturbed by switching on $\phi_\textmd{G}$ (solid line) and relaxes (for $t >0$) before reaching the new equilibrium at $t \to \infty$. The color gradient from blue (small sizes) to red (large sizes) visualizes the magnitude of the sizes of the colloids.}
  \label{fig:fig1}
\end{figure}

We initiate our system at equilibrium and consider two time-dependent protocols. In the first one, the external potential $\phi$ is absent for $t<0$, and is activated for $t>0$ ({\it switch on}) to observe the ensuing relaxation towards the new equilibrium stated (reached in the limit $t\rightarrow \infty$) (cf. Fig.~\ref{fig:fig1} as a representative illustration). In the second protocol we follow the inverse process: the field is already activated for $t<0$, and it is {\it switched off} for $t>0$. Comparison between both procedures will allow to examine whether the system shows a different relaxation dynamics during the switching on and switching off processes.

This bidirectional exploration is conducted for two external potentials, namely gravitational ($\phi_\textmd{G}$) and osmotic ($\phi_\textmd{O}$) potentials: 

(\textit{i}) The gravitational external potential is linear in 'height' $z$ and reads as
\begin{equation}
    \phi_\textmd{G}(z) = Az \quad \text{for} \quad 0 \le z \le L,
    \label{eq:Bz}
\end{equation}
with $\beta A \sigma_0=1$. Note that the gravitational potential is just a function of position $z$, not of size. Still, density inhomogeneities will also affect the compressible particle sizes and their distributions in time and space.

(\textit{ii}) The osmotic potential, $\phi_\textmd{O}(z,\sigma)$ is given by
\begin{equation}
    \phi_\textmd{O}(z,\sigma) = Bz\sigma^3 \quad \text{for} \quad 0 \le z \le L,
    \label{eq:Oz}
\end{equation}
where $\beta B \sigma_0^4=1$. This equation introduces a dependence on particle size through the $\sigma^3$ volume term, aiming to replicate the volumetric effects exerted by osmotic pressure in environments with varying cosolute concentrations with a constant concentration gradient. As such, $\sigma^3$ represents the volume exclusion effect of particles within this gradient.~\cite{Denton2002,Lin2020}

To validate the theoretical predictions of our extended DDFT, they are compared to BD simulations results. Through this comparison, we analyze macroscopic quantities such as the mean position $\langle z(t) \rangle$, mean size $\langle \sigma(t) \rangle$, and the pressure exerted on the hard walls. Moreover, we delve into a parametric study focusing on the dynamics of these responsive systems, governed by the parameter $\alpha$. For $\alpha <1$ the swelling/deswelling of the responsive colloid is slower than the translational diffusion, which means that the change of particle size happens later than the structural relation. The opposite occurs for $\alpha >1$. This dissimilar time relaxation is expected to lead to transient dynamic states, that will be explored in the following section.

Fig.~\ref{fig:fig1} provides a representative illustration of the activation protocol for the particular case of the gravitational external potential. Big and small particles are depicted as red and blue spheres, respectively. The rest of intermediate sizes are represents by a continuous graduation of colors between red and blue. At time $t<0$ (left panel) the responsive system is in the equilibrium state, confined between two planar hard walls. After activation of the external field (central panel) at $t=0$, particles dynamically relax in the new potential field and migrate towards the left to lower the external potential energy. Their sizes are also affected subjected to the changes in the local particle concentration. For $t\rightarrow \infty$ (right panel) the system finally reaches the final equilibrium state in presence of the external potential, with a very different profile and size distribution compared to the original state. 

\subsection{Gravitational potential}

\begin{figure*}[ht!]
    \includegraphics[width=\linewidth]{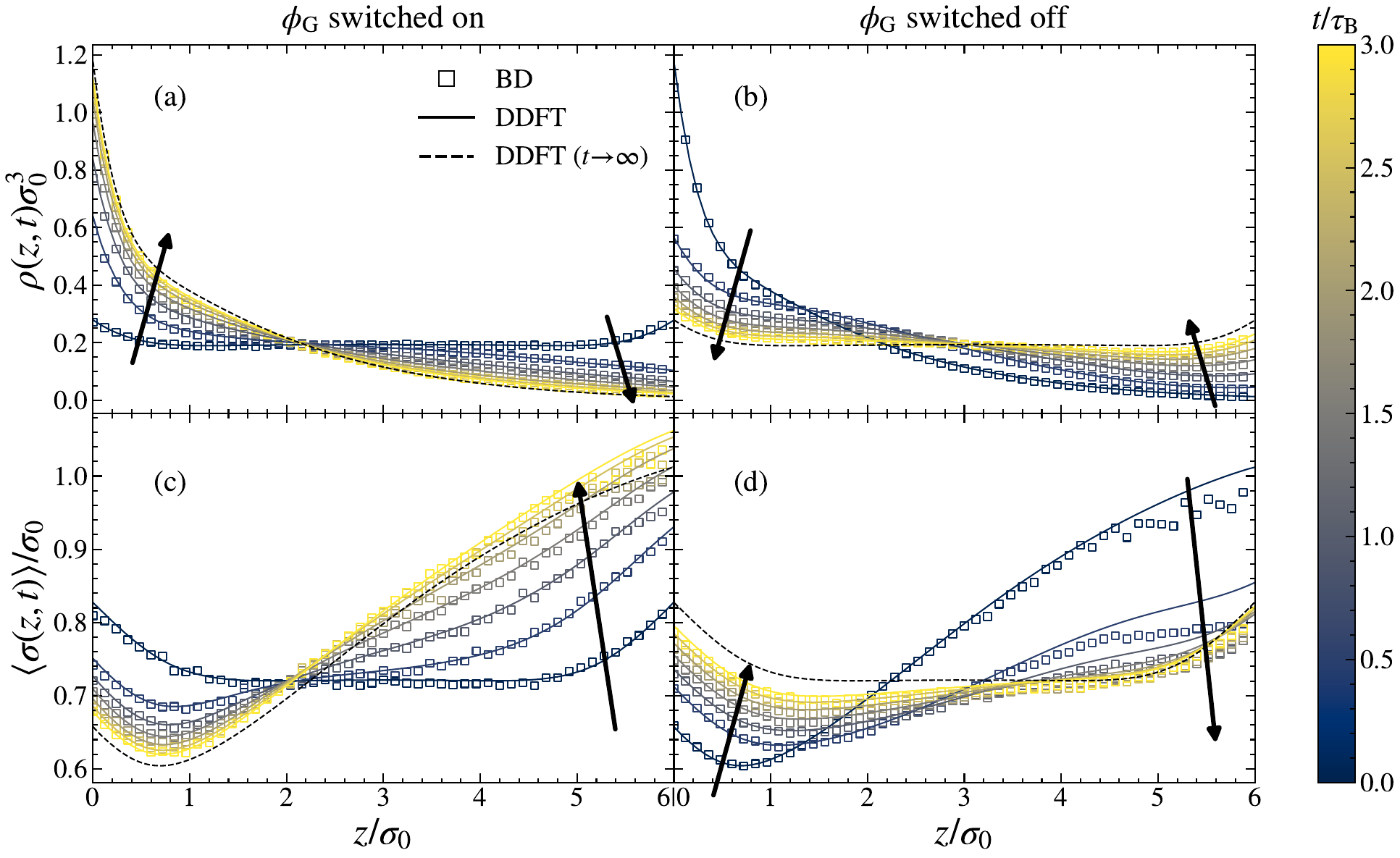}
    \caption{Time evolution of the density $\rho(z;t)$ and size $\langle \sigma(z;t)\rangle$ of the RCs in the gravitational field, $\phi_\textmd{G}(z)$, plotted at different times ranging from $t=0$ (dark blue) to $3\tau_{\rm B}$ (yellow), with a time interval of $\Delta t=0.5\tau_{\rm B}$. Lines represent the theoretical predictions obtained with RC-DDFT, whereas symbols correspond to BD simulations. Panels (a) and (c) show, respectively, the local density ($\rho(z;t)$) and the local mean size ($\left\langle \sigma(z;t) \right\rangle$) obtained when the external potential is switched on for $t>0$. Similarly, panels (b) and (d) show the same quantities, but for the deactivation process ('switched off'). In this case, $\left\langle \sigma(z) \right\rangle$ data of RC-DDFT is multiplied by a factor $1.05$ for a clearer comparison.  All calculations are performed considering $\alpha=0.1$ and a surface density $N/S = 1.2\sigma_0^{-2}$.}
  \label{fig:fig2}
\end{figure*}

We start analyzing the relaxation dynamics of a responsive colloidal system immersed in the gravitational external potential given by Eq.~(\ref{eq:Bz}) under the two protocols (switching on and off) outlined in the preceding section. For this case, we select $\alpha=0.1$ and $N/S = 1.2\sigma_0^{-2}$. Fig.~\ref{fig:fig2} depicts the time evolution of the mean local density ($\rho(z;t)$) and local mean size ($\langle \sigma(z;t) \rangle$) from $t=0$ to $3\tau_{\rm B}$ at $\Delta t =0.5\tau_{\rm B}$ intervals. Panels (a) and (c) show the results following the activation ('switch on') of the gravitational field, while panels (b) and (d) depict the outcomes subsequent to its deactivation ('switch off') . 

The analysis of $\rho(z)$, depicted in panels (a) and (b), reveals a significant alignment between the BD and RC-DDFT methods throughout the dynamic process. These panels also indicate that the system has reached nearly the final equilibrium state already for $t\approx 3\tau_{\rm B}$. Conversely, the evaluation of $\langle \sigma(z;t) \rangle$ in panels (c) and (d) demonstrates consistent conformity between the methods, albeit with notable differences, particularly a consistent disparity of approximately 5\% throughout the dynamic sequence. This deviation in $\langle \sigma(z;t) \rangle$ arises not only in this confined geometry but also in bulk systems (without external potentials and walls). We attribute this discrepancy to the inherent limitations of the mean-field approximation employed in RC-DDFT. To facilitate clearer comparison in the figures, this difference has been rectified in the RC-DDFT data by a scaling factor of $1.05$. Following this adjustment, the temporal evolution described by both techniques closely corresponds.

In the initial stage ($t=0$) of the activation state, depicted in panel (a) of Fig.~\ref{fig:fig2}, we observe a nearly flat density profile, $\rho(z)$, with the exception of regions near the walls where particle adsorption effects become prominent. This accumulation near the hard walls occurs because particles are pushed from the bulk to the walls due to the interparticle repulsive interactions. In addition, the local mean size, $\langle \sigma(z) \rangle$, as shown in Fig.\ref{fig:fig2}(c), exhibits intriguing behavior near the wall. Typically, in the bulk, the mean size is expected to decrease as $\rho$ increases.~\cite{Baul2021} However, near the walls, an increase in $\langle \sigma \rangle$ is observed as $\rho$ increases, a phenomenon also noted in equilibrium configurations.~\cite{moncho2024external} This behavior can be explained by the conditions faced by colloids near a hard boundary. Unlike in the bulk, where colloids are fully surrounded by other particles, being near a wall reduces the number of neighboring particles by half. This reduction in surrounding particles diminishes the overall repulsive forces acting on the colloids, allowing them to expand.

Upon activation of the gravitational field $\phi_\textmd{G}(z)$ at time $t>0$, the system is subjected to a reordering as particles migrate towards the region with lower external field, located on the left side of the slit. This process, depicted in the dynamic sequence in Figs.~\ref{fig:fig2}(a) and (c) is not instantaneous; thus, we plot the density and mean size at various times to capture the evolution. It is interesting to remark the strong change of $\langle \sigma(z) \rangle$ during the first stages of the evolution ($0.5\tau_{\rm B}$), compared to the evolution of $\rho(z;t)$. Given that $\alpha=0.1$ implies that size diffusion should be significantly slower than spatial diffusion, the observed rapid change in size must be necessarily caused by Stokes-type diffusion, $D_\textmd{T} \sim 1/\sigma$: Smaller particles, having a higher diffusion coefficient, move faster towards the left, leaving behind larger particles and thus increasing the average size on the right side of the system. Additionally, a decrease in density favors more expanded states for the particles.

The system keeps evolving, and we illustrate these dynamics only up to $3\tau_{\rm B}$ because the changes become exceedingly slow thereafter. By $10\tau_{\rm B}$, the system reaches a state indistinguishable from the equilibrium state achieved under the gravitational field. In this final equilibrium state, which is also the starting point for the deactivation process (Figs.~\ref{fig:fig2}(b) and (d)), the density $\rho$ decays from the left wall. In an ideal, non-interacting system, this decay would be purely exponential. However, due to the existent repulsive interactions between colloids, the actual final equilibrium density profile departs from this ideal behavior. Excluding the near-wall effects previously discussed, regions with higher particle concentration correspond to smaller mean sizes, indicating compression due to increased density. Interestingly, near the right wall, where $\rho$ is nearly zero, the mean size is maximal and very close to one, suggesting negligible interparticle interactions as the size distribution aligns with the parent distribution, $p(\sigma)$. The complete dynamics of the activation process depicted in Fig.~\ref{fig:fig1} are illustrated through representative snapshots obtained from Brownian Dynamics (BD) simulations, capturing the initial, intermediate, and final states. As depicted, upon activation of $\phi_\textmd{G}(z)$, responsive particles undergo noticeable diffusion towards the left wall. This results in compression, leading to a significant localization effect in particle size when compared to the less compressed region near the right wall.

\begin{figure*}[ht!]
    \includegraphics[width=\linewidth]{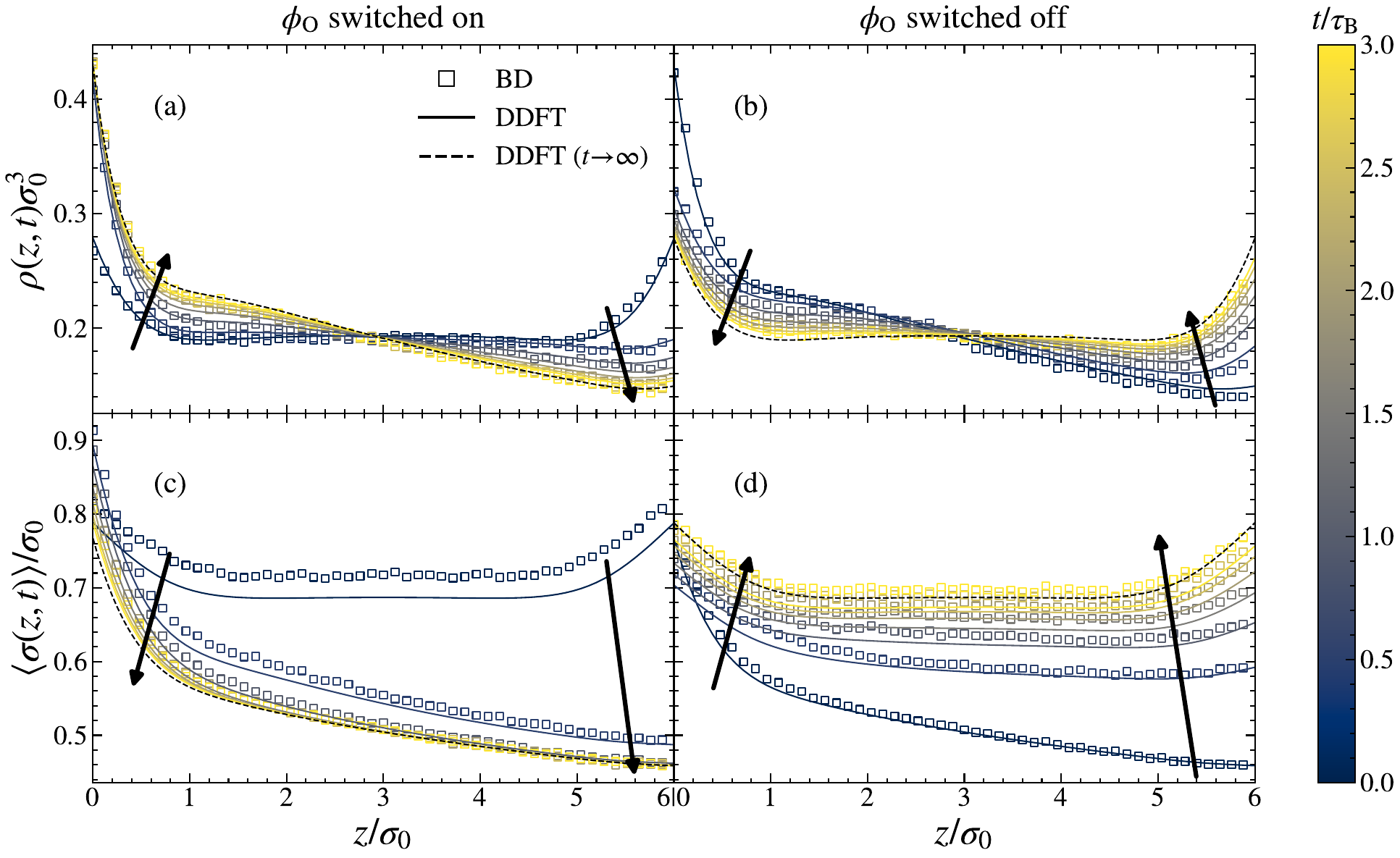}
    \caption{Time evolution of the RCs in the osmotic field, $\phi_\textmd{O}(z,\sigma)$, plotted at different times ranging from $t=0$ (dark blue) to $3\tau_{\rm B}$ (yellow), with a time interval of $\Delta t=0.5\tau_{\rm B}$. Lines represent the theoretical predictions obtained with RC-DDFT, whereas symbols correspond to BD simulations. Panels (a) and (c) show respectively the local density ($\rho(z;t)$) and the local mean size ($\left\langle \sigma(z;t) \right\rangle$) obtained when the external potential is switched on for $t>0$. Similarly, panels (b) and (d) show the same quantities, but for the deactivation process.
    All calculations are performed considering $\alpha=0.1$ and a surface density $N/S = 1.2\sigma_0^{-2}$.}
  \label{fig:fig3}
\end{figure*}

Figs.~\ref{fig:fig2}(b) and (d) show again $\rho(z;t)$ and $\langle \sigma(z;t) \rangle$ for the deactivation process, respectively. In this case, we find that the migration towards the new equilibrium state is not symmetrical compared to the activation process. During activation, particle movement is driven by the external gravitational force, whereas upon deactivation, particle movement is driven by the gradient of concentration, which pushes the particle from the more dense region (left) to the more dilute one (right). This asymmetry results in differing dynamics between activation and deactivation phases, with the gravitational field's application appearing to accelerate the dynamic process as it will be discussed later.

\subsection{Osmotic potential}

Next, we turn to examine the dynamics of the RC fluid when the osmotic potential is activated and deactivated (Eq.~\ref{eq:Oz}). This external potential varies linearly with $z$, akin to gravitational force, but its pronounced $\sigma^3$-dependency gives rise to a markedly distinct qualitative dynamic behavior. Fig.~\ref{fig:fig3}(a) and (c) illustrate the mean particle density and mean size within the planar slit during the activation process, respectively, while plots (b) and (d) delineate the same quantities during deactivation. We maintain $\alpha=0.1$ and $N/S = 1.2\sigma_0^{-2}$.

Upon activation of $\phi_\textmd{O}(z,\sigma)$, particles tend to migrate globally towards the left, mitigating the energy contribution induced by the external potential, expressed as $\int \rho \phi_\textmd{O}dzd\sigma$. Analogous to gravitational effects, this results in a notable increase in particle density near the left wall over time.

However, $\langle \sigma(z;t) \rangle$ exhibits contrasting behavior, with larger particle sizes accompanying high-density regions, as depicted in Fig.~\ref{fig:fig3}(c). This occurs despite the compression near the left wall. The phenomenon can be elucidated by the strong dependence of $\phi_\textmd{O}(z,\sigma)$ on $\sigma$: the energetic penalty for larger colloids near the right wall prompts smaller colloids to prevail in that region. Additionally, $\langle \sigma(z;t) \rangle$ experiences rapid decline in early stages of evolution ($t < 0.5\tau_\textmd{B}$), not attributable to RC shrinkage, given that calculations are conducted with a deswelling diffusion $10$ times smaller than spatial diffusion. This effect stems from the migration of larger colloids towards the left wall, propelled by the applied external force, $f_\textmd{ext}=-d\phi_\textmd{O}/dz=-A\sigma^3$.

Distinct dynamical behavior emerges upon deactivating the osmotic potential. Remarkably, convergence is achieved within $3\tau_{\rm B}$ in both scenarios. However, activating the osmotic potential leads to markedly faster dynamics compared to its deactivation. This acceleration is particularly pronounced when examining the size distribution, with temporal profiles nearly coinciding after just $t\approx \tau_{\rm B}$.

Furthermore, transient behavior is observed upon activation of $\phi_\textmd{O}(z,\sigma)$. Indeed, after activation, there is a rapid migration of colloids towards the left wall, resulting in a significant concentration increase at very short times in this region. Concurrently, the mean particle size of colloids near the left wall initially increases, followed by a decay over longer times. This clearly indicates that particles first move to the left and then decreases their size.

Conversely, deactivating $\phi_\textmd{O}(z,\sigma)$ yields the opposite trend, with $\rho$ and $\sigma$ increasing on the left and decreasing on the right until equilibrium is attained.

In both scenarios (activation and deactivation), the comparison between theoretical predictions from RC-DDFT and BD simulations demonstrates very good agreement.

\subsection{Macroscopic Analysis}

\begin{figure*}[ht!]
    \centering
    \includegraphics[width=\linewidth]{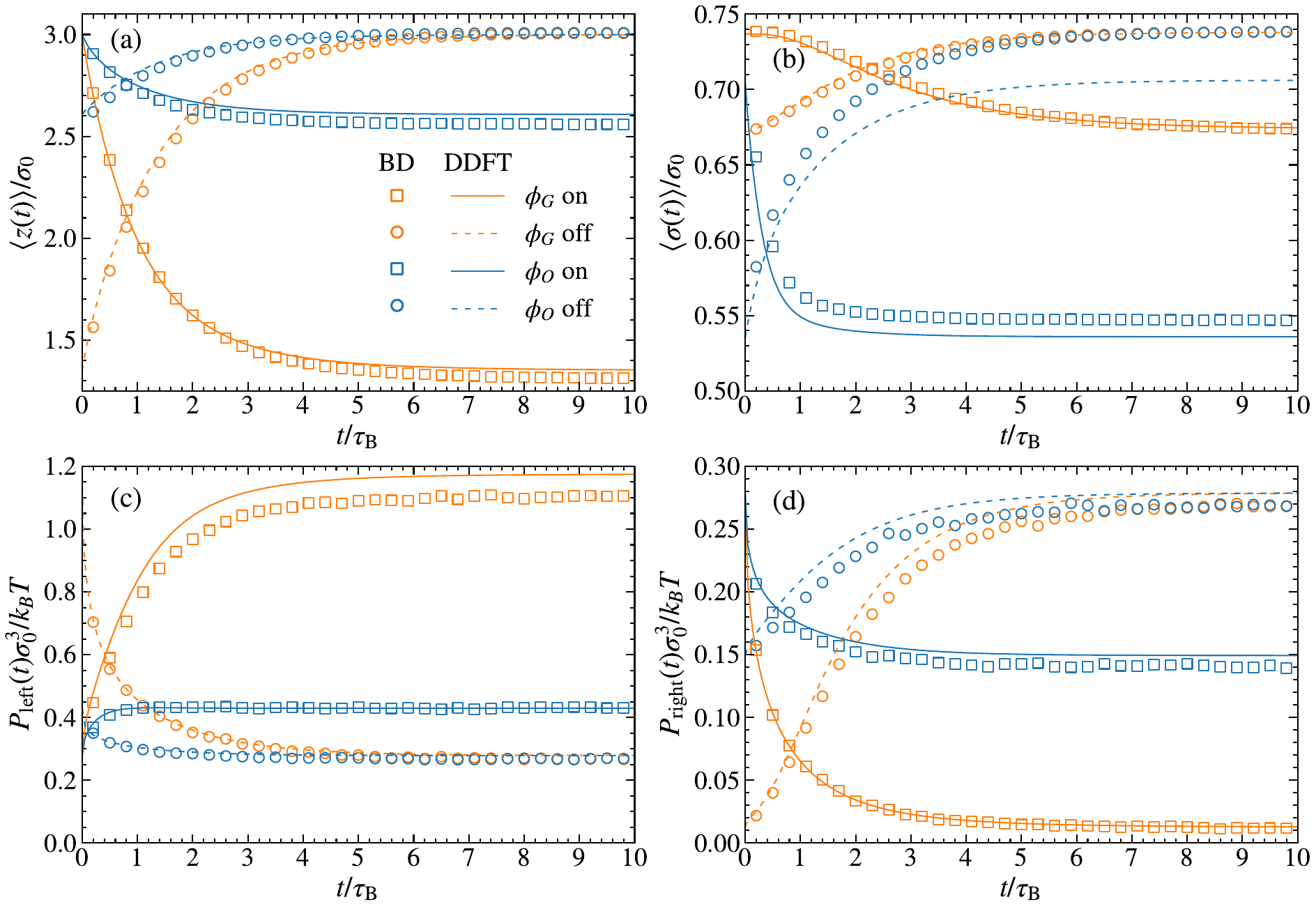}
    \caption{Time evolution of (a) the center of mass position ($\langle z(t) \rangle$), (b) the mean particle size of the system ($\langle \sigma(t) \rangle$, with gravitational RC-DDFT data multiplied by 1.05), (c) the pressure applied by the fluid on the left wall ($P_\textmd{left}(t)$), and (d) on the right wall $P_{right}(t)$, obtained for the activation (solid lines) and deactivation (dashed lines) processes. Orange and blue colors represents the results for the gravitational and osmotic potentials, respectively. Lines denote RC-DDFT predictions, whereas symbols BD simulations. In all cases $\alpha=0.1$ and a surface density $N/S = 1.2\sigma_0^{-2}$.}
    \label{fig:fig4}
\end{figure*}

\begin{figure*}[ht!]
    \centering
    \includegraphics[width = \textwidth]{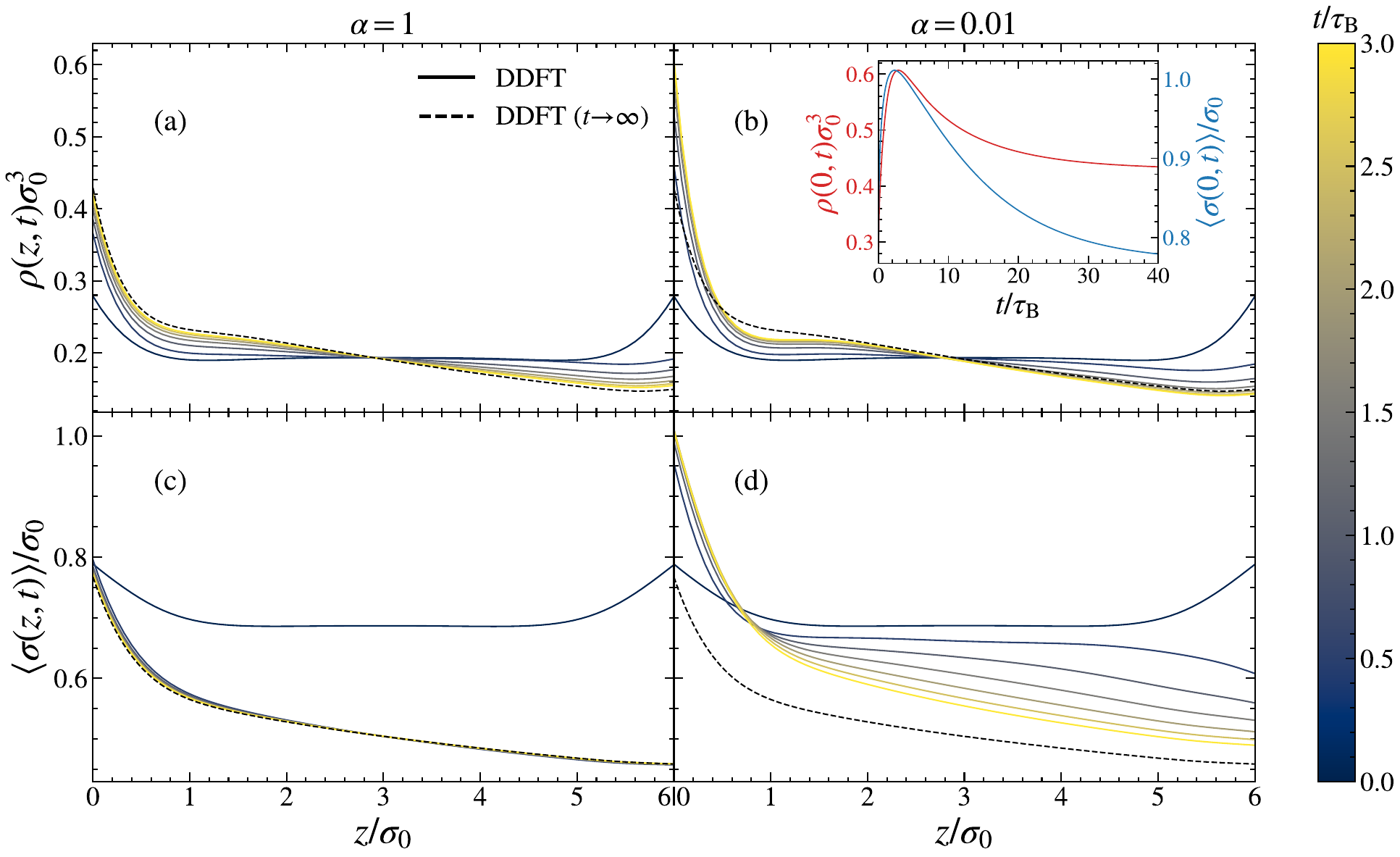}
    \caption{RC-DDFT results for the time evolution of the system structure for the osmotic field, $\phi_\textmd{O}$, being switched on at time $t=0$ for various values of the timescale parameter $\alpha$. (a), (b) Local density, $\rho(z)$, and (c), (d) local mean size, $\left\langle \sigma(z) \right\rangle$, at different times ranging from 0 $\tau_{\rm B}$ (blue) to 3 $\tau_{\rm B}$ (yellow) with a time interval of 0.5 $\tau_{\rm B}$. The surface density is $N/S=1.2\sigma_0^{-2}$. The insert in (b) shows the time evolution of the density $\rho$ and the mean size $\left\langle \sigma \right\rangle$ on the left wall.}
    \label{fig:fig_5}
\end{figure*}

To validate the agreement between RC-DDFT and BD beyond the local density and size profiles, we examine three interesting macroscopic, integrated quantities. These quantities not only offer insights into the global behavior of the system but also facilitate a comparison of the time scales on which dynamical processes occur. The quantities of interest are the center of mass position $\langle z \rangle$ of the suspension (Eq.~(\ref{eq:meanz})), the global mean particle size $\langle \sigma(t) \rangle$ (Eq.~(\ref{eq:meansigma})), and the pressures exerted by the RC fluid on the left and right hard walls, given by $P_\textmd{left}(t)$ and $P_\textmd{right}(t)$ (Eq.~(\ref{eq:pressleft})), respectively. 

The time evolution of these quantities is illustrated in Fig.~\ref{fig:fig4} for $\alpha=0.1$ and $N/S = 1.2\sigma_0^{-2}$. Each panel presents results obtained from RC-DDFT for activated (solid lines) and deactivated (dashed lines) gravitational and osmotic external potentials, distinguished by orange and blue lines, respectively. Corresponding BD simulations are represented by squared and circled symbols. Remarkable congruence between macroscopic quantities derived from both RC-DDFT and BD, not only in shape but also in indicative time scales (the values of DDFT-$\sigma$ under gravitational field has been multiplied again by a factor 1.05).

The analysis of these curves reveals several features of the relaxation process. Primarily, it is evident that relaxation during the activation of the external field is consistently faster than during deactivation. We attribute this behavior to the supplementary driving force induced by the activated external potential, intensifying the motion of confined colloids. Conversely, during the deactivation process, this external force is absent, resulting in colloids relying solely on diffusion for movement, thereby leading to a slower evolution towards equilibrium. 

With our system's inherent two degrees of freedom, we employ a double exponential function to fit these macroscopic quantities, facilitating the extraction of time scales:

\begin{equation}
    \psi(t)=A_1 e^{-t/\tau_1} + A_2 e^{-t/\tau_2} + c,
    \label{eq:fit}
\end{equation}
being $\psi(t)=\{\langle z (t)\rangle, \langle\sigma(t)\rangle, P_L (t), P_L (t)\}$. The relaxation times, $\tau_1$ and $\tau_2$, obtained from these fits, are summarized in Table \ref{tab:relax_times}, presenting a comparison of the time scales predicted by RC-DDFT and observed through BD (between parenthesis) under the scenarios studied.

\begin{table}[]
\caption{\label{tab:relax_times} Relaxation times obtained by fitting Eq.~(\ref{eq:fit}) to mean size,  $\left\langle\sigma(t)\right\rangle$, and mean position, $\left\langle z(t)\right\rangle$ predicted by RC-DDFT during the activation (on) and deactivation (off) of the external potentials, $\phi_\textmd{G}(z)$ and $\phi_\textmd{O}(z,\sigma)$. The corresponding times for BD simulations are shown inside the parenthesis.}
\begin{tabular}{|c|c|ccccccc|} \hline
                           & $\tau$ & $\phi_\textmd{G}$ (on) & &$\phi_\textmd{G}$ (off) & & $\phi_\textmd{O}$ (on) & & $\phi_\textmd{O}$ (off) \\
                            \hline\hline
\multirow{2}{*}{$\langle z\rangle$}  & $\tau_1$    & 0.90 (0.91)  & & 1.38 (1.47)   & & 0.24 (0.25)      & & 1.33 (1.46)               \\
                            & $\tau_2$ & 1.97 (1.95)      &  & 1.38 (1.47)& & 1.23 (1.23)  & & 1.33 (1.46)\\
                            \hline\hline
\multirow{2}{*}{$\langle\sigma\rangle$} & $\tau_1$ & 1.08 (1.15)     & & 0.61 (0.72) & & 0.30 (0.31) & & 0.28 (0.27)        \\
                            & $\tau_2$ & 1.69 (1.83)   & & 1.50 (1.55) & & 1.20 (1.42) & & 1.48 (1.61)                \\
                            \hline
\end{tabular}
\end{table}

In every scenario, we observe two distinct time scales that are similar within each potential and across both RC-DDFT and BD analyses. This similarity suggests that the complex dynamics of the system are influenced by the interaction between translational and size diffusion processes. The differences in the shapes of macroscopic quantities for various potentials, as shown in Fig.~\ref{fig:fig4}, likely arise from the differing impacts of these time scales. The close match between macroscopic quantities obtained from BD simulations and those predicted by RC-DDFT highlights the effectiveness of the RC-DDFT extension in accurately capturing the dynamics of the system. This concordance further supports the use of RC-DDFT for additional studies.

As summarized in Table~\ref{tab:relax_times}, the relaxation time for $\langle z(t)\rangle$ when $\phi_\textmd{O}$ is switched on ($0.24\tau_{\rm B}$) and off ($1.33\tau_{\rm B}$) or the relaxation time for $\langle\sigma(t)\rangle$ when $\phi_\textmd{G}$ is switched on ($1.08\tau_{\rm B}$) and off ($0.61\tau_{\rm B}$) is a clear signature of irreversible relaxation pathway. In addition, the difference in relaxation time for $\langle z(t)\rangle$ when switching on $\phi_\textmd{O}$ and $\phi_\textmd{G}$ reveals that the nature of the perturbation has a important effect on the relaxation process (see Fig.~\ref{fig:fig4}).

\subsection{Competition of time scales: the $\alpha$-study}


Having evaluated the effects of the activation and deactivation process of the external field, we now focus on the assessment of the influence of the time scale ratio, $\alpha=D_\sigma / D_0$, on the non-equilibrium relaxation dynamics of our confined system of soft RCs. As mentioned earlier, $\alpha$ modulates the interplay between size diffusion and translational diffusion. For such a study, we select the osmotic external potential and consider only the activation process. Fig.~\ref{fig:fig_5} shows the time evolution of $\rho(z;t)$ and $\langle \sigma(z;t) \rangle$ for $\alpha=1$ (plots (a) and (c)) and $\alpha=0.01$ (plots (b) and (d)), representing two limiting dynamic condition for which the $\sigma$-diffusion is very fast and very slow compared to the translational diffusion. The particle surface concentration is fixed in all cases to $N/S= 1.2\sigma_0^{-2}$. Considering the close alignment between RC-DDFT and BD observed in previous sections, we only depict the results obtained with RC-DDFT.  

For $\alpha=1$, the particle size rapidly adjusts to the new environmental conditions during the initial relaxation stage, and the system approaches full structural equilibrium within $3\tau_{\rm B}$. Decreasing $\alpha$ to $0.01$ results in a slower response of particle size to the activation of the external potential. In this regime, our findings reveal the presence of a transient dynamic state: initially, particles accumulate on the left wall for $t<3\tau_\textmd{B}$, followed by a subsequent decrease in concentration towards the final equilibrium state for $t>3\tau_\textmd{B}$. This dynamic reentrance is clearly depicted in the inset of Fig.~\ref{fig:fig_5}(b), illustrating the time dependence of the particle density in contact with the left hard wall, $\rho(z=0,t)$, which exhibits a maximum at $t\approx 3\tau_B$. A similar trend is observed for the mean size of colloids in contact with the left wall ($\langle \sigma(z=0,t) \rangle$), initially increasing with time before eventually decreasing again (see again the inset of Fig.~\ref{fig:fig_5}(b)).

It is important to highlight that this transient dynamic state observed at intermediate times arises directly from the disparity in time scales between translation and swelling, evident for $\alpha=0.01$ and absent for $\alpha=1$. Initially, colloids diffuse towards the left under the external field without altering their size, which evolves $100$ times slower. The observed increase in $\langle \sigma(z=0,t) \rangle$ is primarily attributed to the selective motion of larger colloids driven by the applied external force $f_\textmd{ext}=-d\phi_\textmd{O}/dz=-A\sigma^3$. Over longer timescales, compression effects lead to a gradual (slower) reduction in particle size, consequently inducing a decrease in $\rho(z=0,t)$ until the new equilibrium state is attained. In essence, this observation signifies a transition from translationally driven to size-driven relaxation dynamics as $\alpha$ decreases from $1$ to $ 0.01$, a characteristic exclusive to responsive colloids.

\begin{figure}
  \centering
    \includegraphics[width=\linewidth]{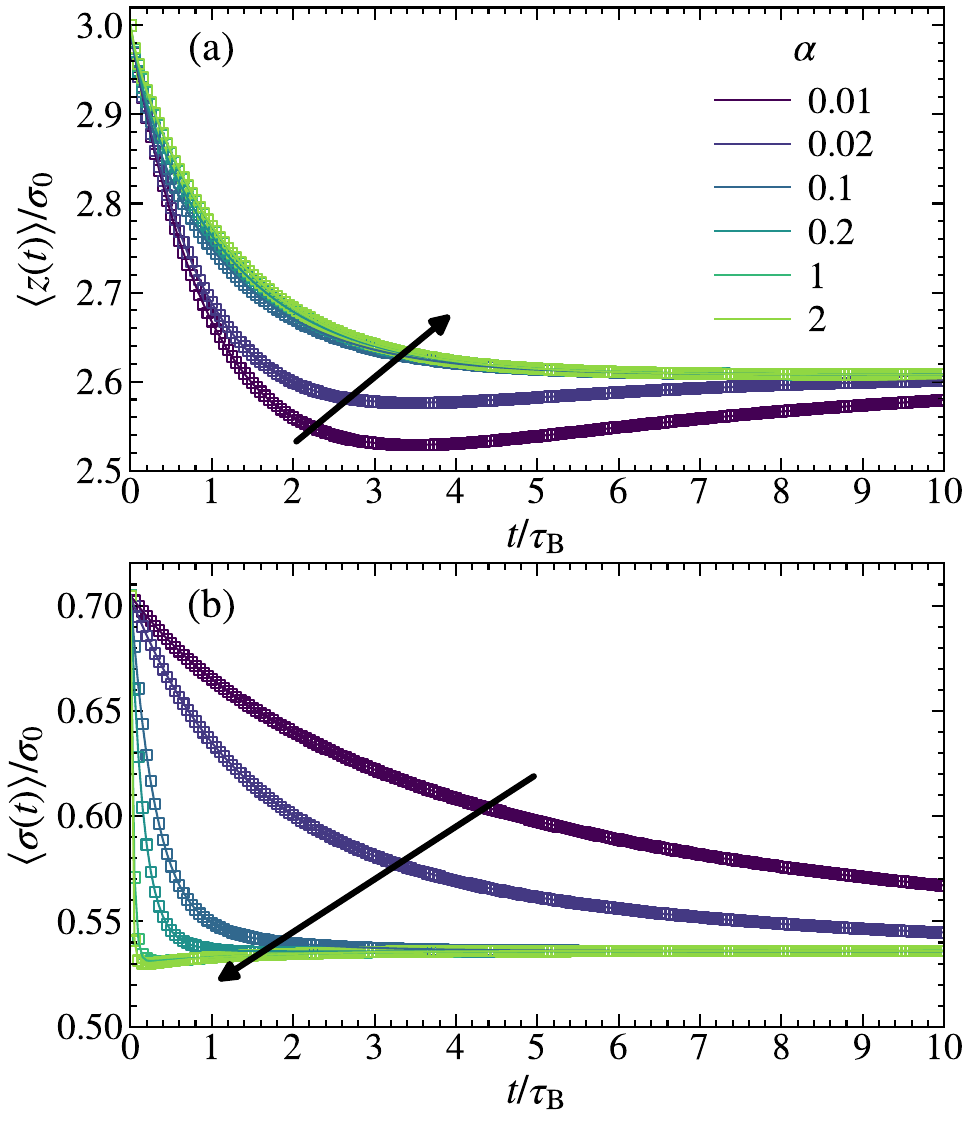}
\caption{Time evolution of (a) the center of mass position ($\left\langle z(t)\right\rangle$) and (b) the mean size of the colloids ($\left\langle\sigma(t)\right\rangle$) predicted by RC-DDFT (symbols) during the activation of the osmotic external potential, obtained for different values of $\alpha$ from $0.01$ to $2$. Solid lines are the result of the fitting of the data using Eq.~(\ref{eq:fit}). In all cases $N/S = 1.2\sigma_0^{-2}$.}
      \label{fig:fig6}
\end{figure}

This rich interplay between structural and size relaxation is further illustrated in Fig.~\ref{fig:fig6}(a) and (b), which explore, respectively, the impact of $\alpha$ on the center of mass position ($\langle z(t) \rangle$) and on the mean particle size ($\langle \sigma(t) \rangle$) across a wide range of $\alpha$ values from $0.01$ to $2$. In cases where $\alpha<0.04$, a non-monotonic curve with a pronounced minimum appears in $\langle z(t) \rangle$, signifying that responsive colloids initially migrate towards the left wall due to the applied external force before subsequently reversing direction due to gradual size reduction. Conversely, for $\alpha>0.25$, the opposing trend is found: the particle size undergoes rapid reduction attributed to the compression induced by the osmotic field during the initial stages of evolution, yet the colloids have not traversed the necessary distance to achieve structural relaxation. For longer times, the progressive redistribution of colloids towards the left wall through spatial diffusion allows $\langle \sigma(t) \rangle$ to exhibit a slight increase, driven by the swelling of colloids near $z=0$.

\begin{figure}
  \centering    \includegraphics[width=\linewidth]{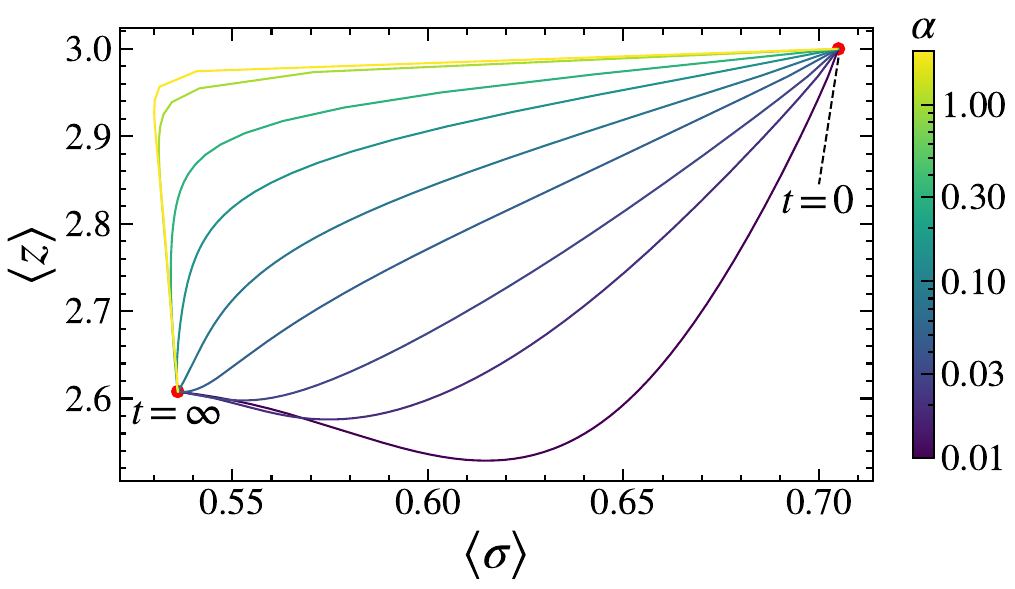}
\caption{Transition pathways: Center of mass location $\langle z \rangle$ versus the mean particle size $\langle \sigma \rangle$ of the system during the relaxation process after activation of the osmotic external potential, plotted for different values of $\alpha$. The red point on the right-top represents the initial equilibrium state ($t=0$), whereas the one located on the left-bottom corresponds to the new equilibrium state ($t\rightarrow \infty$). Results are obtained using RC-DDFT with $N/S = 1.2\sigma_0^{-2}$.}
  \label{fig:fig7}
\end{figure}

This transient behavior observed for dissimilar values of $D_\textmd{T}$ and $D_\sigma$ becomes evident in Fig.~\ref{fig:fig7}, where $\langle z(t) \rangle $ is plotted against $\langle \sigma(t) \rangle$ for different values of $\alpha$, from $0.01$ to $3$. Indeed, for $\alpha=0.01$ we find that $\langle z(t) \rangle$ shows a dynamic reentrance induced by the slow $\sigma$-response, leading to a minimum in the curve. Conversely, for $\alpha=3$. the curve depicts a shoulder on the left, indicating the reentrance of the center of mass location caused by the slow translational diffusion.  These phenomena controlled by $\alpha$ demonstrate the dynamic competition between translational and size diffusion. At very low or high $\alpha$ values, the dynamics of mean size and position appear closely coupled, depicting a scenario where one aspect of the system's behavior must "wait" for the other to stabilize.

\begin{figure*}
  \centering
    \includegraphics[width=0.45\linewidth]{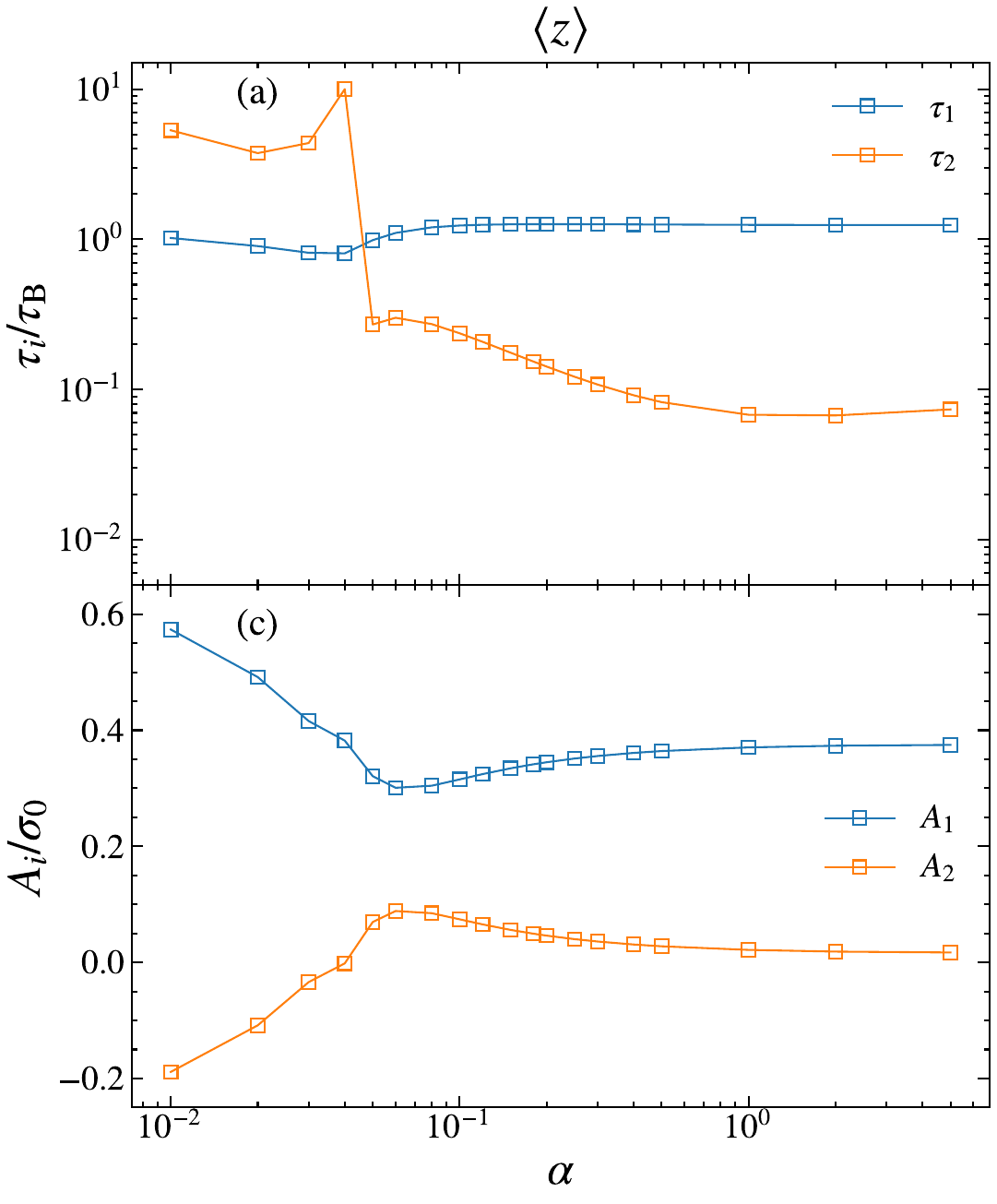}
    \includegraphics[width=0.45\linewidth]{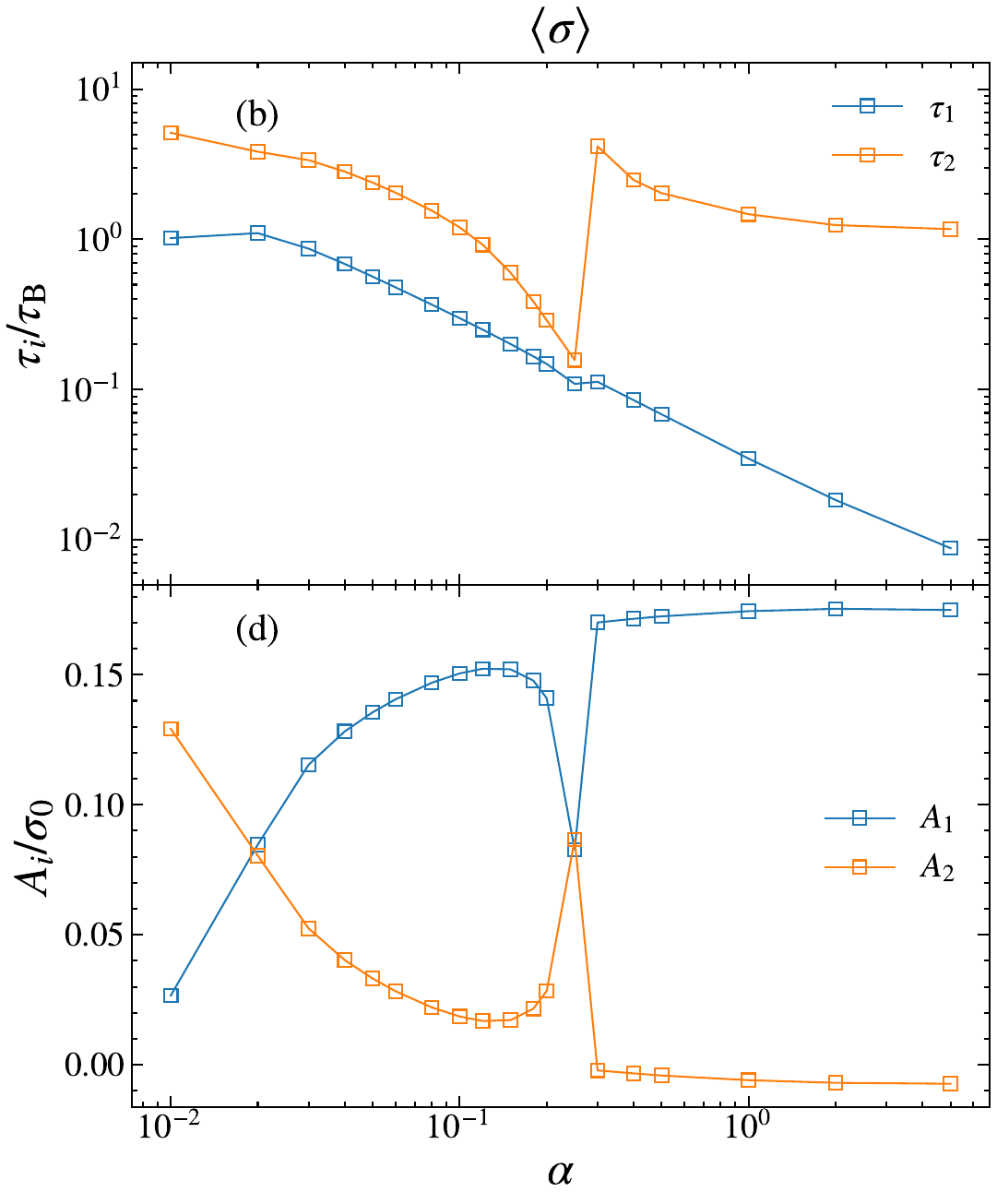}
\caption{Relaxation times, $\tau_i$, and prefactors, $A_i$, as a function of $\alpha$. They are obtained by fitting $\left\langle z(t)\right\rangle$ (plots (a) and (c)) and$\left\langle\sigma(t)\right\rangle$ (plots (b) and (d)) predicted by RC-DDFT to Eq.~(\ref{eq:fit}), during the activation process of the osmotic external potential, for $N/S = 1.2\sigma_0^{-2}$. The data corresponds to the structural time evolution shown in Figs.~\ref{fig:fig6}.}
  \label{fig:fig8}
\end{figure*}

The fitting of mean size and mean position according to the double exponential function given by Eq.~(\ref{eq:fit}) yields relaxation times and prefactors as functions of $\alpha$, as shown in Figure~\ref{fig:fig8}. This analysis confirms the presence of two distinct time scales. 
Examining the mean position, as depicted in Fig.~\ref{fig:fig8}(a), enables us to classify $\tau_1$ as the translational relaxation time and $\tau_2$ as the size relaxation time. Notably, $\tau_1$ remains relatively unchanged across different values of $\alpha$. In scenarios where $\alpha$ is high, the prefactor $A_1$ is always significantly larger then $A_2$ (please not that in this regime $A_2$ is very close to $0$.), indicating that size adjusts swiftly, and the mean position is predominantly governed by translational relaxation. Conversely, at lower $\alpha$ values, the significance of $A_2$ elevates, taking negative values. In this regime the translational relaxation transpires swiftly following a change in size, suggesting that the mean position, $\langle z \rangle$, is driven by size adjustments. Taking a look to the relaxation times of the mean size, depicted in  Fig.~\ref{fig:fig8}(b), we can conversely identify $\tau_1$ as the size relaxation, since it follows a power law with $\alpha$, and $\tau_2$ as the translational or density relaxation. For high $\alpha$ the relaxation in size is mainly dominated by size diffusion, since $A_1 \gg A_2$.


The conspicuous discontinuities observed in Figs. \ref{fig:fig8}(a) and (b) might be interpreted, from a physical perspective, as indicative of a behavioral transition in the dynamical relaxation. Specifically, the domain characterized by high $\alpha$ values can be conceptualized as a conventional liquid-like phase where translational relaxation predominates as the chief mechanism. Conversely, the domain with low $\alpha$ values mirrors a more glassy/amorphous phase, with size relaxation emerging as the primary mechanism and a linked localized cage relaxation. This delineation hints at a fundamental shift in the system's behavior, transitioning from one dominated by translational dynamics to one where size relaxations play a more critical role. 

\section{Concluding remarks}
\label{sec:conclusions}

In this work, we have developed and combined a dynamic density-functional theory (RC-DDFT) framework and Brownian dynamics (BD) simulations to investigate the full time-dependent non-equilibrium relaxation dynamics of confined systems of soft responsive colloids (RCs) after activation/deactivation of external potentials. In contrast to conventional models of soft colloids, in our model the size dynamics of the colloids is explicitly resolved and the influence of its relaxation behavior and timescale on the full system could be explored for the first time.  The results showed a complex interplay between the translational diffusion and particle swelling/shrinking, leading to interesting non-equilibrium structuring as well as reentrant transient states when the typical spatial and size relaxation times scales are very different. We also demonstrated that the modification of intrinsic timescales of the system lead to tuneable macroscopic relaxation times and pathways in systems with many relaxing degrees of freedom. The excellent agreement between DDFT and BD showed that DDFT can also be faithfully used to study the nonequilibrium behavior of soft, weakly correlated colloids with additional internal degrees of freedom, if the systems are not too slaved by dissipative mechanisms.~\cite{schmidt2022power}  Hence, our study constitutes a big step forward for the modeling and description of the full nonequilibrium structuring of soft complex fluids. 

In future, it would be interesting to extend the RC-DDFT method to responsive (intrinsically polydisperse) systems of stiffer systems, e.g., microgels modeled by Hertizan potentials.~\cite{Urich2016, Rovigatti2019} For these systems, we expect to find marked oscillations of the density profiles that will become strongly affected by particle stiffness. Also the dynamical properties can lead to the formation of interesting transient, e.g., highly structured while full nonequilibrium regions that finally relax when the colloids adapt their size to the new environmental conditions. These transients may have interesting new properties. Finally, one could also envision to add more internal degrees of freedom to the colloids with a more complex hierarchy of timescales, or even add internal chemical activity to the colloids, e.g., as used in catalytically active nanoreactors where the bistable size response is crucial for complex self-dynamics.~\cite{Milster2023}

\begin{acknowledgments}
    This work was supported by the Deutsche Forschungsgemeinschaft (DFG) via the Research Unit FOR 5099 “Reducing complexity of nonequilibrium systems” and Project No. 430195928. The authors also acknowledge support by the state of Baden-Württemberg through bwHPC and the DFG through grant no INST 39/963-1 FUGG (bwForCluster NEMO). Further support was provided by grant PID2022-136540NB-I00 funded by MCIN/AEI/10.13039/501100011033, ERDF \textit{A way of making Europe}, the Spanish Ministerio de Ciencia e Innovación, Programa Estatal de Investigación Científica, Técnica y de Innovación 2021–2023 (Project No. PID2022-136540NB-00) and program Visiting Scholars funded by the Plan Propio of the University of Granada (Project No. PPVS2018-08). J.L.M. thanks the Ph.D. student fellowship (FPU21/03568) supported by the Spanish \textit{Ministerio de Universidades}.
\end{acknowledgments}


%
%

%


\bibliography{RC_bimodal}

\end{document}